\pdfoutput=1
\documentclass[11pt,letterpaper]{article}

\usepackage[english]{babel}
\usepackage[utf8]{inputenc}
\usepackage[T1]{fontenc}
\usepackage{lmodern}
\usepackage[letterpaper,textwidth=6.0in,textheight=8.6in,centering]{geometry}
\usepackage{amsmath,amsthm,amssymb,mathtools}
\usepackage{booktabs}
\usepackage{microtype}
\usepackage{enumitem}
\usepackage{array}
\usepackage{tikz}
\usetikzlibrary{arrows.meta,positioning,calc}
\usepackage{algorithm}
\usepackage{algpseudocode}
\usepackage[colorlinks=true,linkcolor=blue!50!black,citecolor=blue!50!black,urlcolor=blue!50!black]{hyperref}

\hypersetup{
  pdftitle={Matching Markets: Equilibrium Computation, Indivisible Goods, and Positional Externalities},
  pdfauthor={Bruce Changlong Xu, Andrew Yang, and Ivan Villa-Renteria},
  pdfsubject={Computer Science and Game Theory (cs.GT)}
}

\theoremstyle{plain}
\newtheorem{theorem}{Theorem}[section]
\newtheorem{lemma}[theorem]{Lemma}
\newtheorem{proposition}[theorem]{Proposition}
\newtheorem{corollary}[theorem]{Corollary}

\theoremstyle{definition}
\newtheorem{definition}[theorem]{Definition}
\newtheorem{example}[theorem]{Example}

\theoremstyle{remark}
\newtheorem{remark}[theorem]{Remark}

\newcommand{\NP}{\mathsf{NP}}
\newcommand{\coNP}{\mathsf{coNP}}
\newcommand{\Pclass}{\mathsf{P}}
\newcommand{\PPAD}{\mathsf{PPAD}}
\newcommand{\PoA}{\operatorname{PoA}}
\newcommand{\SW}{\operatorname{SW}}
\newcommand{\prob}[1]{\textnormal{\textsc{#1}}}

\title{\textbf{Matching Markets: Equilibrium Computation,\\[0.15em]
Indivisible Goods, and Positional Externalities}}

\author{
  Bruce Changlong Xu \qquad Andrew Yang \qquad Ivan Villa-Renteria \\[0.35em]
  {\normalsize Stanford University}
}

\date{}

\begin{document}
\maketitle

\begin{abstract}
\noindent Matching markets provide a mathematical framework for allocating scarce resources among strategic agents with heterogeneous preferences. While two-sided and one-sided matching without money rely on ordinal stability and Pareto efficiency, Fisher and Arrow--Debreu markets mediate allocation through budgets and market-clearing prices. In this paper, we study the convex-analytic foundations, computational complexity, and external-utility extensions of Fisher matching markets. First, we formulate fractional resource allocation and linear Fisher markets via Fenchel conjugate duality and the Karush--Kuhn--Tucker (KKT) conditions of the Eisenberg--Gale convex program, showing how budget-weighted Nash social welfare characterizes competitive equilibria across linear and homothetic utility families. Second, we review topological existence proofs via Sperner's lemma alongside polynomial-time combinatorial algorithms for divisible markets, including the Devanur--Papadimitriou--Saberi--Vazirani (DPSV) primal-dual max-flow algorithm and Orlin's strongly polynomial scaling algorithm. Third, we contrast the tractability of divisible markets with the $\NP$-hardness and $\PPAD$-completeness of indivisible Fisher markets and Competitive Equilibrium from Equal Incomes (CEEI), comparing dynamic-programming $\varepsilon$-approximate equilibria, \prob{CEEI-FRAC}, and Budish's Approximate CEEI mechanism. Finally, we introduce and analyze Fisher markets with \emph{positional (snob-effect) consumption externalities}, in which buyer $i$'s utility for good $j$ scales with her relative consumption share $(x_{ij}/\sum_{k} x_{kj})^{\alpha_i}$ for sensitivity $\alpha_i > 0$. We prove that positional utility summands are convex on the nonnegative orthant, implying that utilitarian social welfare is maximized at exclusive corner allocations where each good is assigned to a single buyer, whereas decentralized price-taking behavior admits a unique budget-proportional equilibrium in single-good snob markets and incurs a Price of Anarchy of $n^\alpha$ among $n$ symmetric buyers.

\vspace{0.65em}
\end{abstract}

\section{Introduction}
\label{sec:intro}

Resource allocation and matching problems lie at the intersection of microeconomic theory, combinatorial optimization, and algorithmic game theory. In settings where monetary transfers are prohibited or repugnant---such as assigning medical residents to hospitals, students to public schools, on-campus housing to undergraduates, or incompatible patient--donor pairs in kidney exchange---allocations are governed by ordinal preferences, stability, and strategyproofness \cite{GaleShapley1962,ShapleyScarf1974,RothSotomayor1990,RothSonmezUnver2004,SonmezUnver2011}. When agents instead possess monetary budgets or artificial currency endowments, prices serve as a decentralized coordination device: each item is assigned a unit price, each buyer demands a utility-maximizing bundle within her budget, and a \emph{competitive equilibrium} (or Walrasian market-clearing equilibrium) is reached when aggregate demand equals supply for every positively priced good \cite{ArrowDebreu1954,BrainardScarf2005}.

The canonical model of price-mediated allocation with fixed budgets is the \emph{Fisher market}, introduced by Irving Fisher in 1891 using a hydraulic analogy \cite{BrainardScarf2005} and analyzed convex-analytically by Eisenberg and Gale \cite{EisenbergGale1959,Eisenberg1961}. In a linear Fisher market, $n$ buyers with budgets $B_1, \dots, B_n > 0$ compete for $m$ divisible goods with supplies $C_1, \dots, C_m > 0$, and buyer $i$'s utility is a linear function $u_i(\mathbf{x}_i) = \sum_{j=1}^m v_{ij} x_{ij}$ of her allocation $\mathbf{x}_i = (x_{i1}, \dots, x_{im}) \in \mathbb{R}_{\ge 0}^m$. Despite the apparent nonlinearity of the market-clearing conditions---which couple individual utility maximization at unknown prices $\mathbf{p}$ with global supply exhaustion---Eisenberg and Gale \cite{EisenbergGale1959} discovered that competitive equilibria of a linear Fisher market coincide with the optimal primal-dual solutions of a single convex program maximizing the budget-weighted geometric mean of buyers' utilities, $\sum_{i=1}^n B_i \log u_i$.

From a computational standpoint, Fisher markets exhibit three striking phenomena that motivate this paper:
\begin{enumerate}[label=(\arabic*),topsep=3pt,itemsep=3pt]
  \item \textbf{From nonconstructive fixed points to strongly polynomial combinatorial algorithms:}
  Classical existence proofs for competitive equilibrium invoke Kakutani's or Brouwer's fixed-point theorem, or equivalently Sperner's combinatorial lemma on a triangulated price simplex \cite{ArrowDebreu1954,Scarf1967,Su1999}. While path-following simplicial algorithms require exponential time in the worst case, the KKT structure of the Eisenberg--Gale program enabled Devanur, Papadimitriou, Saberi, and Vazirani (DPSV) \cite{Devanur2008,Vazirani2007} to design a primal-dual ascending-price algorithm based on parametric max-flow in an equality subgraph, later refined by Orlin \cite{Orlin2010} into a strongly polynomial $O((n+m)^4 \log(n+m))$ scaling algorithm.

  \item \textbf{Sharp complexity transitions under indivisibility and equal incomes:}
  When goods are indivisible ($\mathbf{x}_i \in \mathbb{Z}_{\ge 0}^m$), competitive equilibria need not exist, and deciding whether an exact equilibrium exists becomes $\NP$-hard even for three buyers with linear utilities \cite{DengPapadimitriouSafra2003}. In the fair-division specialization where all buyers receive an equal budget---Competitive Equilibrium from Equal Incomes (CEEI)---exact existence remains weakly $\NP$-hard for two buyers and strongly $\NP$-hard for $n$ buyers and $3n$ items \cite{Aziz2015,BouveretLemaitre2016,Branzei2015}, while computing Budish's Approximate CEEI (A-CEEI) in combinatorial assignment markets \cite{Budish2011} is $\PPAD$-complete \cite{OthmanPapadimitriouRubinstein2016}.

  \item \textbf{Breakdown of welfare-equilibrium equivalence under positional externalities:}
  Standard Fisher market theory assumes that buyer $i$'s utility $u_i(\mathbf{x}_i)$ depends solely on her own bundle $\mathbf{x}_i$. In many markets for status goods, luxury items, congestible resources, and priority access, however, agents exhibit what Leibenstein \cite{Leibenstein1950} termed the \emph{snob effect}: the utility buyer $i$ derives from good $j$ diminishes when other buyers also consume a large share of good $j$. Once utilities depend on relative consumption shares $x_{ij} / \sum_{k=1}^n x_{kj}$, the concavity assumptions underlying the Eisenberg--Gale program and the First Welfare Theorem no longer hold.
\end{enumerate}

\paragraph{Organization and contributions.}
Section~\ref{sec:preliminaries} develops the mathematical preliminaries of fractional matching, linear Fisher markets, Fenchel conjugate duality, and the KKT characterization of the Eisenberg--Gale convex program. Section~\ref{sec:divisible} presents the Sperner-labeling existence proof for divisible Fisher markets alongside the DPSV primal-dual max-flow algorithm and Orlin's strongly polynomial algorithm. Section~\ref{sec:indivisible} surveys the complexity hierarchy of indivisible Fisher markets, $\varepsilon$-approximate equilibria, \prob{CEEI}, \prob{CEEI-FRAC}, and Budish's \prob{A-CEEI} mechanism. In Section~\ref{sec:snob}, we study Fisher markets with positional (snob-effect) consumption externalities $U_i(\mathbf{x}) = \sum_{j=1}^m v_{ij} x_{ij} (x_{ij} / \sum_{k=1}^n x_{kj})^{\alpha_i}$. We prove that for all $\alpha_i > 0$, each positional utility term is convex on the nonnegative orthant, so utilitarian social welfare is maximized at exclusive corner allocations where each good is assigned to a single buyer (Theorem~\ref{thm:snob-welfare}). We then characterize decentralized competitive equilibria in single-good snob markets, proving that budget-proportional sharing is the unique market-clearing equilibrium and establishing an exact Price of Anarchy of $n^\alpha$ across $n$ symmetric buyers (Theorem~\ref{thm:snob-equilibrium}). Section~\ref{sec:conclusion} concludes with open algorithmic questions.

\section{Preliminaries: Fisher Markets and Eisenberg--Gale Duality}
\label{sec:preliminaries}

Throughout this paper, let $\mathcal{N} = \{1, 2, \dots, n\}$ denote a set of $n$ buyers (agents) and $\mathcal{M} = \{1, 2, \dots, m\}$ denote a set of $m$ goods (items), with available supply $C_j > 0$ for each good $j \in \mathcal{M}$.

\subsection{Fractional Matching and Assignment}

Before introducing budgets and prices, consider the baseline fractional assignment problem in which each buyer $i \in \mathcal{N}$ requires a unit bundle of goods ($\sum_{j \in \mathcal{M}} x_{ij} = 1$, assuming $\sum_{j \in \mathcal{M}} C_j \ge n$). An allocation is a matrix $\mathbf{x} = (x_{ij})_{i \in \mathcal{N},\, j \in \mathcal{M}} \in \mathbb{R}^{n \times m}$ satisfying the linear feasibility constraints:
\begin{align}
  \sum_{j \in \mathcal{M}} x_{ij} &= 1,   &&\forall i \in \mathcal{N}, \label{eq:match-unit}\\
  \sum_{i \in \mathcal{N}} x_{ij} &\le C_j, &&\forall j \in \mathcal{M}, \label{eq:match-supply}\\
  x_{ij} &\ge 0,                          &&\forall i \in \mathcal{N},\; \forall j \in \mathcal{M}. \label{eq:match-nonneg}
\end{align}
Constraint~\eqref{eq:match-unit} requires that each buyer $i \in \mathcal{N}$ receive one total unit of goods (possibly assembled from fractional shares across multiple items); constraint~\eqref{eq:match-supply} enforces that the aggregate allocation of each good $j \in \mathcal{M}$ not exceed its supply $C_j$; and constraint~\eqref{eq:match-nonneg} enforces nonnegativity.

\begin{example}
Suppose $\mathcal{N} = \{1, 2\}$ and $\mathcal{M} = \{1, 2, 3\}$ with supplies $C_1 = 2$, $C_2 = 0.5$, and $C_3 = 0.5$. Then the allocation
\[
  (x_{11}, x_{12}, x_{13}) = (0.5, 0.4, 0.1), \qquad
  (x_{21}, x_{22}, x_{23}) = (0.7, 0.1, 0.2)
\]
satisfies~\eqref{eq:match-unit}--\eqref{eq:match-nonneg}, since each row sums to $1$ and the column sums $(1.2, 0.5, 0.3)$ are bounded by $(2, 0.5, 0.5)$.
\end{example}

When each buyer $i \in \mathcal{N}$ derives value $v_{ij} \ge 0$ per unit of good $j \in \mathcal{M}$, a benevolent social planner might seek to maximize utilitarian welfare $\sum_{i \in \mathcal{N}} \sum_{j \in \mathcal{M}} v_{ij} x_{ij}$, or to maximize the Nash product $\prod_{i \in \mathcal{N}} (\sum_{j \in \mathcal{M}} v_{ij} x_{ij})$ to balance efficiency and fairness. In a decentralized market, however, buyers are not constrained to consume a fixed physical quantity $\sum_j x_{ij} = 1$; instead, each buyer is constrained by a monetary budget and chooses her own bundle in response to item prices.

\subsection{Linear Fisher Markets and Competitive Equilibrium}
\label{subsec:fisher-def}

In a \emph{Fisher market}, each buyer $i \in \mathcal{N}$ is endowed with a positive monetary budget $B_i > 0$ (which has no intrinsic utility outside the market), and each good $j \in \mathcal{M}$ has supply $C_j > 0$. By rescaling the units in which each good $j$ is measured (replacing $x_{ij}$ by $x_{ij}/C_j$ and $v_{ij}$ by $C_j v_{ij}$), we may normalize supplies to $C_j = 1$ for all $j \in \mathcal{M}$ whenever convenient; we retain general $C_j > 0$ when discussing indivisible goods.

A utility function $u_i: \mathbb{R}_{\ge 0}^m \to \mathbb{R}_{\ge 0}$ for buyer $i$ satisfies $u_i(\mathbf{0}) = 0$ and is monotonically non-decreasing in each coordinate (free disposal). The fundamental utility classes studied in Fisher market theory belong to the Constant Elasticity of Substitution (CES) family:
\begin{equation}
  u_i(\mathbf{x}_i) = \Biggl(\sum_{j \in \mathcal{M}} v_{ij} x_{ij}^\rho\Biggr)^{\!1/\rho}, \qquad \rho \in (-\infty, 1] \setminus \{0\},
  \label{eq:ces-utility}
\end{equation}
which recovers:
\begin{itemize}[topsep=2pt,itemsep=2pt]
  \item \textbf{Linear utilities ($\rho = 1$, perfect substitutes):} $u_i(\mathbf{x}_i) = \sum_{j \in \mathcal{M}} v_{ij} x_{ij}$, where $v_{ij} \ge 0$;
  \item \textbf{Cobb--Douglas utilities ($\rho \to 0$):} $u_i(\mathbf{x}_i) = \prod_{j \in \mathcal{M}} x_{ij}^{v_{ij}}$ with $\sum_{j \in \mathcal{M}} v_{ij} = 1$;
  \item \textbf{Leontief utilities ($\rho \to -\infty$, perfect complements):} $u_i(\mathbf{x}_i) = \min_{j:\, v_{ij} > 0} (x_{ij} / v_{ij})$.
\end{itemize}

In the linear Fisher market, given a nonnegative price vector $\mathbf{p} = (p_1, \dots, p_m) \in \mathbb{R}_{\ge 0}^m$ (where $p_j$ is the price per unit of good $j$), each buyer $i \in \mathcal{N}$ solves the individual linear program:
\begin{equation}
\begin{aligned}
  &\text{maximize}_{\mathbf{x}_i \in \mathbb{R}^m} &&\sum_{j \in \mathcal{M}} v_{ij} x_{ij} \\
  &\text{subject to} &&\sum_{j \in \mathcal{M}} p_j x_{ij} \le B_i, \\
  &                  &&x_{ij} \ge 0, \qquad \forall j \in \mathcal{M}.
\end{aligned}
\label{eq:buyer-lp}
\end{equation}
Let $\mathcal{D}_i(\mathbf{p}) \subseteq \mathbb{R}_{\ge 0}^m$ denote the set of optimal solutions to~\eqref{eq:buyer-lp}, called the \emph{demand set} of buyer $i$ at prices $\mathbf{p}$. When $p_j > 0$ for all $j \in \mathcal{M}$ and $v_{ij} > 0$ for at least one $j$, buyer $i$'s \emph{maximum bang-per-buck} (maximum utility per dollar) at prices $\mathbf{p}$ is
\begin{equation}
  \alpha_i(\mathbf{p}) := \max_{j \in \mathcal{M}} \frac{v_{ij}}{p_j}.
  \label{eq:bang-per-buck}
\end{equation}
A bundle $\mathbf{x}_i \in \mathbb{R}_{\ge 0}^m$ belongs to $\mathcal{D}_i(\mathbf{p})$ if and only if $\sum_{j \in \mathcal{M}} p_j x_{ij} = B_i$ and $x_{ij} > 0$ implies $v_{ij}/p_j = \alpha_i(\mathbf{p})$ (buyer $i$ spends her entire budget exclusively on goods maximizing her bang-per-buck).

\begin{definition}[Competitive Equilibrium]
\label{def:ce}
A pair $(\mathbf{x}^*, \mathbf{p}^*)$ consisting of an allocation $\mathbf{x}^* = (x_{ij}^*) \in \mathbb{R}_{\ge 0}^{n \times m}$ and a price vector $\mathbf{p}^* = (p_1^*, \dots, p_m^*) \in \mathbb{R}_{\ge 0}^m$ is a \emph{competitive equilibrium} (CE), or \emph{market-clearing equilibrium}, of the Fisher market if:
\begin{enumerate}[label=(\roman*),topsep=2pt,itemsep=2pt]
  \item \textbf{Buyer optimality:} For every buyer $i \in \mathcal{N}$, $\mathbf{x}_i^* \in \mathcal{D}_i(\mathbf{p}^*)$ (every buyer receives a utility-maximizing affordable bundle at prices $\mathbf{p}^*$, exhausting her budget $B_i$ whenever $\max_j v_{ij} > 0$);
  \item \textbf{Market clearing:} For every good $j \in \mathcal{M}$, $\sum_{i \in \mathcal{N}} x_{ij}^* \le C_j$, and if $p_j^* > 0$, then $\sum_{i \in \mathcal{N}} x_{ij}^* = C_j$ (every positively priced good is completely sold, and any undersold good has price zero).
\end{enumerate}
\end{definition}

Throughout our analysis of divisible linear Fisher markets, we make the standard non-degeneracy assumption that every good has at least one potential buyer ($\forall j \in \mathcal{M},\, \exists i \in \mathcal{N}$ with $v_{ij} > 0$) and every buyer desires at least one good ($\forall i \in \mathcal{N},\, \exists j \in \mathcal{M}$ with $v_{ij} > 0$). Under this condition, any equilibrium price vector satisfies $p_j^* > 0$ for all $j \in \mathcal{M}$ and $\sum_{j \in \mathcal{M}} C_j p_j^* = \sum_{i \in \mathcal{N}} B_i$.

\subsection{Fenchel Conjugate Duality and KKT Optimality Conditions}
\label{subsec:kkt}

To connect market equilibria with convex optimization, we briefly recall Fenchel conjugate duality and the Karush--Kuhn--Tucker (KKT) conditions \cite{BoydVandenberghe2004,JainVazirani2010}. For a convex function $f: \mathbb{R}^d \to (-\infty, +\infty]$, its \emph{convex conjugate} (or Fenchel--Legendre transform) $f^*: \mathbb{R}^d \to (-\infty, +\infty]$ is defined by
\begin{equation}
  f^*(\boldsymbol{\mu}) = \sup_{\mathbf{x} \in \mathrm{dom}(f)} \bigl\{\boldsymbol{\mu}^\top \mathbf{x} - f(\mathbf{x})\bigr\}.
  \label{eq:fenchel}
\end{equation}
Economically, if $f(\mathbf{x})$ represents the cost of producing a vector of quantities $\mathbf{x}$ and $\boldsymbol{\mu}$ represents unit market prices, then $f^*(\boldsymbol{\mu})$ is the maximum profit attainable at prices $\boldsymbol{\mu}$. Geometrically, $-f^*(\boldsymbol{\mu})$ is the vertical intercept of the supporting hyperplane to the epigraph of $f$ with slope $\boldsymbol{\mu}$.

Given a matrix $A = (a_{kj}) \in \mathbb{R}^{m \times d}$, vectors $\mathbf{b} \in \mathbb{R}^m$ and $\mathbf{c} \in \mathbb{R}^d$, and a differentiable convex function $f: \mathbb{R}^d \to \mathbb{R}$, consider the primal concave maximization problem and its Fenchel--Lagrangian dual:
\begin{equation}
\begin{aligned}
  &\textbf{Primal:} &&\max_{\mathbf{x} \ge \mathbf{0}} \;\; \sum_{j=1}^d c_j x_j - f(\mathbf{x})
  &&\text{s.t.}\quad \sum_{j=1}^d a_{kj} x_j \le b_k \;\;\forall k \in \{1,\dots,m\}, \\
  &\textbf{Dual:}   &&\min_{\boldsymbol{\lambda} \ge \mathbf{0},\, \boldsymbol{\mu}} \;\; \sum_{k=1}^m b_k \lambda_k + f^*(\boldsymbol{\mu})
  &&\text{s.t.}\quad \sum_{k=1}^m a_{kj} \lambda_k \ge c_j - \mu_j \;\;\forall j \in \{1,\dots,d\}.
\end{aligned}
\label{eq:fenchel-primal-dual}
\end{equation}

More generally, consider a convex optimization problem in standard form:
\begin{equation}
  \min_{\mathbf{x} \in \mathbb{R}^d} \; f_0(\mathbf{x}) \qquad \text{subject to} \quad g_k(\mathbf{x}) \le 0 \;\;\forall k \in \{1,\dots,m\}, \quad h_\ell(\mathbf{x}) = 0 \;\;\forall \ell \in \{1,\dots,p\},
  \label{eq:std-convex}
\end{equation}
where $f_0, g_1, \dots, g_m$ are differentiable convex functions and $h_1, \dots, h_p$ are affine. Introducing dual multipliers $\boldsymbol{\alpha} \in \mathbb{R}_{\ge 0}^m$ and $\boldsymbol{\nu} \in \mathbb{R}^p$, the Lagrangian is $\mathcal{L}(\mathbf{x}, \boldsymbol{\alpha}, \boldsymbol{\nu}) = f_0(\mathbf{x}) + \sum_{k=1}^m \alpha_k g_k(\mathbf{x}) + \sum_{\ell=1}^p \nu_\ell h_\ell(\mathbf{x})$. Whenever Slater's constraint qualification holds (for instance, when all constraints are linear and feasible), a point $\mathbf{x}^*$ and dual multipliers $(\boldsymbol{\alpha}^*, \boldsymbol{\nu}^*)$ are primal-dual optimal if and only if they satisfy the four \emph{Karush--Kuhn--Tucker (KKT) conditions}:
\begin{enumerate}[label=\textbf{(KKT-\arabic*)},leftmargin=*,topsep=2pt,itemsep=2pt]
  \item \textbf{Primal feasibility:} $g_k(\mathbf{x}^*) \le 0$ for all $k \in \{1,\dots,m\}$, and $h_\ell(\mathbf{x}^*) = 0$ for all $\ell \in \{1,\dots,p\}$.
  \item \textbf{Dual feasibility:} $\alpha_k^* \ge 0$ for all $k \in \{1,\dots,m\}$.
  \item \textbf{Complementary slackness:} $\alpha_k^* g_k(\mathbf{x}^*) = 0$ for all $k \in \{1,\dots,m\}$ (an inequality constraint is either active with $g_k(\mathbf{x}^*) = 0$, or inactive with multiplier $\alpha_k^* = 0$).
  \item \textbf{Stationarity of the Lagrangian:}
  \[
    \nabla f_0(\mathbf{x}^*) + \sum_{k=1}^m \alpha_k^* \nabla g_k(\mathbf{x}^*) + \sum_{\ell=1}^p \nu_\ell^* \nabla h_\ell(\mathbf{x}^*) = \mathbf{0}.
  \]
\end{enumerate}

\subsection{The Eisenberg--Gale Convex Program}
\label{subsec:eg-program}

We now show how the competitive equilibria of a linear Fisher market with unit supplies ($C_j = 1$ for all $j \in \mathcal{M}$) are captured by the Eisenberg--Gale convex program \cite{EisenbergGale1959,Eisenberg1961,JainVazirani2010}. Two natural invariance axioms motivate the objective function:
\begin{enumerate}[label=(\arabic*),topsep=2pt,itemsep=2pt]
  \item \textbf{Utility scale invariance:} Rescaling any buyer $i$'s valuation vector $(v_{i1}, \dots, v_{im}) \mapsto c_i (v_{i1}, \dots, v_{im})$ by a positive constant $c_i > 0$ does not alter her ordinal preferences or demand set $\mathcal{D}_i(\mathbf{p})$, and therefore must leave the equilibrium allocation unchanged.
  \item \textbf{Buyer-splitting invariance:} If buyer $i$ with budget $B_i = B_i^{(1)} + B_i^{(2)}$ is split into two buyers $i^{(1)}, i^{(2)}$ with identical valuations $v_{i^{(1)}j} = v_{i^{(2)}j} = v_{ij}$ and budgets $B_i^{(1)}, B_i^{(2)} > 0$, the sum of their equilibrium allocations must equal an equilibrium allocation for $i$.
\end{enumerate}
The budget-weighted geometric mean of utilities (Nash social welfare) satisfies both axioms:
\[
  \max_{\mathbf{x} \ge \mathbf{0}} \Biggl(\prod_{i \in \mathcal{N}} u_i^{B_i}\Biggr)^{\!\frac{1}{\sum_{i \in \mathcal{N}} B_i}}.
\]
Because the logarithm is strictly increasing and $\sum_{i \in \mathcal{N}} B_i > 0$ is constant, maximizing the budget-weighted geometric mean is equivalent to maximizing
\[
  \arg\max_{\mathbf{x}} \prod_{i \in \mathcal{N}} u_i^{B_i}
  = \arg\max_{\mathbf{x}} \log\Biggl(\prod_{i \in \mathcal{N}} u_i^{B_i}\Biggr)
  = \arg\max_{\mathbf{x}} \sum_{i \in \mathcal{N}} B_i \log u_i.
\]
This yields the \emph{Eisenberg--Gale (EG) convex program}:
\begin{equation}
\begin{aligned}
  &\text{maximize}_{\mathbf{x},\, \mathbf{u}} &&\sum_{i \in \mathcal{N}} B_i \log u_i \\
  &\text{subject to} &&u_i = \sum_{j \in \mathcal{M}} v_{ij} x_{ij}, &&\forall i \in \mathcal{N}, \\
  &                  &&\sum_{i \in \mathcal{N}} x_{ij} \le 1,       &&\forall j \in \mathcal{M}, \\
  &                  &&x_{ij} \ge 0,                                 &&\forall i \in \mathcal{N},\; \forall j \in \mathcal{M}.
\end{aligned}
\label{eq:eg-program}
\end{equation}

\begin{theorem}[Eisenberg--Gale Equivalence and Equilibrium Properties \cite{EisenbergGale1959,Vazirani2007}]
\label{thm:eg-kkt}
Assume that for every good $j \in \mathcal{M}$ there is at least one buyer $i \in \mathcal{N}$ with $v_{ij} > 0$, and for every buyer $i \in \mathcal{N}$ there is at least one good $j \in \mathcal{M}$ with $v_{ij} > 0$. Let $p_j \ge 0$ denote the KKT dual multiplier on the supply constraint $\sum_{i \in \mathcal{N}} x_{ij} \le 1$ in~\eqref{eq:eg-program}. Then:
\begin{enumerate}[label=(\alph*),topsep=2pt,itemsep=2pt]
  \item A feasible allocation $\mathbf{x}^*$ and dual vector $\mathbf{p}^*$ are primal-dual optimal for~\eqref{eq:eg-program} if and only if they satisfy the KKT conditions:
  \begin{enumerate}[label=(\roman*),topsep=1pt,itemsep=1pt]
    \item $\forall j \in \mathcal{M}:\; p_j^* \ge 0$;
    \item $\forall j \in \mathcal{M}:\; p_j^* > 0 \implies \sum_{i \in \mathcal{N}} x_{ij}^* = 1$;
    \item $\forall i \in \mathcal{N},\; \forall j \in \mathcal{M}:\; \dfrac{v_{ij}}{p_j^*} \le \dfrac{u_i^*}{B_i}$, where $u_i^* = \sum_{k \in \mathcal{M}} v_{ik} x_{ik}^*$;
    \item $\forall i \in \mathcal{N},\; \forall j \in \mathcal{M}:\; x_{ij}^* > 0 \implies \dfrac{v_{ij}}{p_j^*} = \dfrac{u_i^*}{B_i}$.
  \end{enumerate}
  \item Any primal-dual optimal solution $(\mathbf{x}^*, \mathbf{p}^*)$ of~\eqref{eq:eg-program} is a competitive equilibrium of the linear Fisher market, and conversely every competitive equilibrium is primal-dual optimal for~\eqref{eq:eg-program}.
  \item Equilibrium utilities $\mathbf{u}^* = (u_1^*, \dots, u_n^*)$ and equilibrium prices $\mathbf{p}^* = (p_1^*, \dots, p_m^*)$ exist and are \emph{unique}, the set of equilibrium allocations is a convex polytope, and if all $v_{ij}$ and $B_i$ are rational, there exists a rational equilibrium allocation and the unique equilibrium prices are rational with polynomial bit complexity.
\end{enumerate}
\end{theorem}
\begin{proof}
Substituting $u_i = \sum_{j \in \mathcal{M}} v_{ij} x_{ij}$ into the objective and associating dual multipliers $p_j \ge 0$ with $\sum_{i \in \mathcal{N}} x_{ij} - 1 \le 0$ and $\mu_{ij} \ge 0$ with $-x_{ij} \le 0$, the Lagrangian of~\eqref{eq:eg-program} (written in maximization form) is
\[
  \mathcal{L}(\mathbf{x}, \mathbf{p}, \boldsymbol{\mu})
  = \sum_{i \in \mathcal{N}} B_i \log\!\Biggl(\sum_{j \in \mathcal{M}} v_{ij} x_{ij}\Biggr)
  - \sum_{j \in \mathcal{M}} p_j \Biggl(\sum_{i \in \mathcal{N}} x_{ij} - 1\Biggr)
  + \sum_{i \in \mathcal{N}} \sum_{j \in \mathcal{M}} \mu_{ij} x_{ij}.
\]
Because the feasible region is compact and there exists a strictly positive allocation $\mathbf{x} > \mathbf{0}$ achieving $u_i > 0$ for all $i \in \mathcal{N}$, the objective is bounded above and attains its maximum at some $\mathbf{x}^*$ with $u_i^* > 0$ for all $i$. Differentiating $\mathcal{L}$ with respect to $x_{ij}$ gives the stationarity condition:
\[
  \frac{\partial \mathcal{L}}{\partial x_{ij}} = \frac{B_i v_{ij}}{u_i^*} - p_j^* + \mu_{ij}^* = 0
  \quad\Longleftrightarrow\quad
  p_j^* - \frac{B_i v_{ij}}{u_i^*} = \mu_{ij}^* \ge 0,
\]
with complementary slackness $\mu_{ij}^* x_{ij}^* = 0$ and $p_j^*(\sum_{i \in \mathcal{N}} x_{ij}^* - 1) = 0$. Since each good $j$ has some buyer $i$ with $v_{ij} > 0$, we must have $p_j^* \ge B_i v_{ij} / u_i^* > 0$, so $p_j^* > 0$ and $\sum_{i \in \mathcal{N}} x_{ij}^* = 1$ for every good $j \in \mathcal{M}$. Dividing $p_j^* \ge B_i v_{ij} / u_i^*$ by $p_j^* B_i / u_i^*$ yields conditions (iii) and (iv), proving part~(a).

To verify part~(b), observe from (iii) and (iv) that for every buyer $i \in \mathcal{N}$, $u_i^*/B_i = \max_{j \in \mathcal{M}} (v_{ij}/p_j^*) = \alpha_i(\mathbf{p}^*)$, so buyer $i$ only receives goods $j$ that maximize her bang-per-buck $v_{ij}/p_j^*$. Multiplying $p_j^* x_{ij}^* = (B_i / u_i^*) v_{ij} x_{ij}^*$ and summing over $j \in \mathcal{M}$ gives
\[
  \sum_{j \in \mathcal{M}} p_j^* x_{ij}^*
  = \frac{B_i}{u_i^*} \sum_{j \in \mathcal{M}} v_{ij} x_{ij}^*
  = \frac{B_i}{u_i^*} u_i^* = B_i.
\]
Hence each buyer $i$ exhausts her budget on maximum bang-per-buck goods ($\mathbf{x}_i^* \in \mathcal{D}_i(\mathbf{p}^*)$), and by (ii) every good is completely sold ($\sum_{i \in \mathcal{N}} x_{ij}^* = 1$). Thus $(\mathbf{x}^*, \mathbf{p}^*)$ is a competitive equilibrium, and the steps reverse verbatim.

For part~(c), strict concavity of $u_i \mapsto B_i \log u_i$ implies that $u_i^*$ takes the same value at every optimal solution of~\eqref{eq:eg-program}. In any equilibrium, each good $j \in \mathcal{M}$ has $x_{ij}^* > 0$ for at least one buyer $i$, so condition~(iv) pins down $p_j^* = B_i v_{ij} / u_i^*$ uniquely. Rationality follows from the combinatorial characterization of tight sets in Section~\ref{subsec:dpsv}.
\end{proof}

Beyond linear utilities, Jain and Vazirani \cite{JainVazirani2010} defined the general class of \emph{Eisenberg--Gale markets} as all resource allocation markets whose equilibria are captured by maximizing $\sum_{i \in \mathcal{N}} B_i \log u_i$ over a linear utility-allocation polytope $\mathcal{P}$. This family includes markets with Leontief utilities \cite{CodenottiVaradarajan2004}, homogeneous utilities of degree one \cite{Eisenberg1961}, Kelly capacity-allocation network flow markets \cite{JainVazirani2010}, and homothetic quasi-concave utilities with production \cite{JainVaziraniYe2005}.

\section{Divisible Fisher Markets: Existence and Polynomial-Time Algorithms}
\label{sec:divisible}

\subsection{Topological Existence via Sperner's Lemma}
\label{subsec:sperner}

Whereas Theorem~\ref{thm:eg-kkt} establishes existence of equilibrium for linear and homogeneous utilities via convex programming, general continuous utility functions do not always admit a convex potential. Instead, existence of competitive equilibrium is established topologically via Sperner's lemma \cite{Scarf1967,Su1999}.

Recall that a $d$-simplex $\Delta^d = \operatorname{conv}\{\mathbf{e}_1, \dots, \mathbf{e}_{d+1}\} \subset \mathbb{R}^{d+1}$ is the convex hull of $d + 1$ affinely independent vertices. For any subset of $k + 1$ vertices, their convex hull is a $k$-face; a $(d-1)$-face spanned by $d$ vertices (obtained by setting one coordinate $p_j = 0$) is called a \emph{facet} $F_j$ opposite vertex $\mathbf{e}_j$. A \emph{simplicial subdivision} (or triangulation) $\mathcal{T}$ of $\Delta^d$ is a finite collection of smaller $d$-simplices (\emph{elementary simplices}) whose union is $\Delta^d$ such that any two intersect in a common face or not at all.

\begin{definition}[Sperner Labeling]
\label{def:sperner}
Let $\mathcal{T}$ be a triangulation of the $d$-simplex $\Delta^d$ with vertex set $V(\mathcal{T})$. A labeling $\ell: V(\mathcal{T}) \to \{1, 2, \dots, d + 1\}$ is a \emph{Sperner labeling} if:
\begin{enumerate}[label=(\roman*),topsep=2pt,itemsep=2pt]
  \item the $d+1$ principal vertices of $\Delta^d$ receive distinct labels: $\ell(\mathbf{e}_j) = j$ for each $j \in \{1, \dots, d+1\}$;
  \item for every vertex $\mathbf{v} \in V(\mathcal{T})$ lying on the boundary of $\Delta^d$, if $\mathbf{v}$ lies on facet $F_j$ (so its $j$-th barycentric coordinate is $v_j = 0$), then $\ell(\mathbf{v}) \neq j$ (equivalently, $\ell(\mathbf{v}) \in \operatorname{supp}(\mathbf{v}) = \{k : v_k > 0\}$).
\end{enumerate}
An elementary $d$-simplex $\sigma \in \mathcal{T}$ is called \emph{fully labeled} (or \emph{panchromatic}) if its $d+1$ vertices carry all $d+1$ distinct labels $\{1, 2, \dots, d+1\}$.
\end{definition}

\begin{figure}[t]
\centering
\begin{tikzpicture}[
  scale=1.15,
  vtx/.style={circle, draw=black!90, fill=white, line width=0.6pt, inner sep=1.4pt, font=\scriptsize},
  pan/.style={circle, draw=black, fill=black!12, line width=0.8pt, inner sep=1.4pt, font=\scriptsize\bfseries}
]
  \coordinate (v00) at (0.0, 0.0);
  \coordinate (v10) at (-0.8, -1.3856);
  \coordinate (v11) at (0.8, -1.3856);
  \coordinate (v20) at (-1.6, -2.7712);
  \coordinate (v21) at (0.0, -2.7712);
  \coordinate (v22) at (1.6, -2.7712);
  \coordinate (v30) at (-2.4, -4.1568);
  \coordinate (v31) at (-0.8, -4.1568);
  \coordinate (v32) at (0.8, -4.1568);
  \coordinate (v33) at (2.4, -4.1568);

  \fill[black!10] (v10) -- (v21) -- (v11) -- cycle;
  \fill[black!10] (v20) -- (v31) -- (v21) -- cycle;
  \fill[black!10] (v21) -- (v31) -- (v32) -- cycle;

  \draw[line width=0.65pt, black!85] (v00) -- (v30) -- (v33) -- cycle;
  \draw[line width=0.5pt, black!75] (v10) -- (v11);
  \draw[line width=0.5pt, black!75] (v20) -- (v22);
  \draw[line width=0.5pt, black!75] (v10) -- (v32);
  \draw[line width=0.5pt, black!75] (v20) -- (v31);
  \draw[line width=0.5pt, black!75] (v11) -- (v31);
  \draw[line width=0.5pt, black!75] (v22) -- (v32);

  \node[vtx] at (v00) {$1$};
  \node[pan] at (v10) {$2$};
  \node[pan] at (v11) {$3$};
  \node[pan] at (v20) {$1$};
  \node[pan] at (v21) {$1$};
  \node[vtx] at (v22) {$1$};
  \node[vtx] at (v30) {$2$};
  \node[pan] at (v31) {$3$};
  \node[pan] at (v32) {$2$};
  \node[vtx] at (v33) {$3$};

  \node[above=3pt of v00, font=\footnotesize] {$\mathbf{e}_1 = (1,0,0)$};
  \node[below left=2pt of v30, font=\footnotesize] {$\mathbf{e}_2 = (0,1,0)$};
  \node[below right=2pt of v33, font=\footnotesize] {$\mathbf{e}_3 = (0,0,1)$};
\end{tikzpicture}
\caption{A Sperner-labeled triangulation of the $2$-simplex $\Delta^2$. Vertices on the left edge ($p_3 = 0$) carry labels in $\{1,2\}$, vertices on the right edge ($p_2 = 0$) carry labels in $\{1,3\}$, and vertices on the bottom edge ($p_1 = 0$) carry labels in $\{2,3\}$. Shaded elementary triangles carry all three labels $\{1,2,3\}$; in accordance with Sperner's lemma, their count ($3$) is odd.}
\label{fig:sperner}
\end{figure}

\begin{lemma}[Sperner's Lemma \cite{Su1999}]
\label{lem:sperner}
Every Sperner-labeled triangulation $\mathcal{T}$ of a $d$-simplex $\Delta^d$ contains an odd number of fully labeled elementary $d$-simplices. In particular, there exists at least one fully labeled elementary $d$-simplex.
\end{lemma}

Figure~\ref{fig:sperner} illustrates Lemma~\ref{lem:sperner} for $d = 2$. Using Sperner's lemma, we obtain a topological proof that competitive equilibria exist in divisible Fisher markets.

\begin{proposition}[Existence of Competitive Equilibrium via Sperner's Lemma \cite{Scarf1967,Su1999}]
\label{prop:sperner-ce}
Consider a divisible Fisher market with $n$ buyers $\mathcal{N} = \{1,\dots,n\}$ and $m$ goods $\mathcal{M} = \{1,\dots,m\}$ with normalized supplies $C_j = 1$ and normalized total budget $\sum_{i=1}^n B_i = 1$. Suppose that for each buyer $i \in \mathcal{N}$, her demand $\mathbf{x}_i(\mathbf{p}) = (x_{i1}(\mathbf{p}), \dots, x_{im}(\mathbf{p}))$ is a continuous function on the interior of the price simplex
\[
  \Delta^{m-1} = \Biggl\{\mathbf{p} = (p_1, \dots, p_m) \in \mathbb{R}_{\ge 0}^m \;\Big|\; \sum_{j=1}^m p_j = 1\Biggr\}
\]
satisfying Walras' law $\sum_{j=1}^m p_j x_{ij}(\mathbf{p}) = B_i$, and that as $p_j \to 0^+$, aggregate demand $z_j(\mathbf{p}) = \sum_{i=1}^n x_{ij}(\mathbf{p})$ exceeds $1$ for some free good ($p_k = 0 \implies z_k(\mathbf{p}) > 1$). Then there exists a market-clearing price vector $\mathbf{p}^* \in \Delta^{m-1}$ such that $z_j(\mathbf{p}^*) = 1$ for all $j \in \mathcal{M}$.
\end{proposition}
\begin{proof}
For any price vector $\mathbf{p} \in \Delta^{m-1}$, summing Walras' law $\sum_{j=1}^m p_j x_{ij}(\mathbf{p}) = B_i$ across all buyers $i \in \mathcal{N}$ yields
\begin{equation}
  \sum_{j=1}^m p_j z_j(\mathbf{p}) = \sum_{i=1}^n B_i = 1 = \sum_{j=1}^m p_j.
  \label{eq:walras-aggregate}
\end{equation}
Since $\mathbf{p} \in \Delta^{m-1}$ is a probability distribution over $\mathcal{M}$ with $\sum_{j=1}^m p_j = 1$, equation~\eqref{eq:walras-aggregate} implies that at least one good $j \in \operatorname{supp}(\mathbf{p})$ with $p_j > 0$ must satisfy $z_j(\mathbf{p}) \le 1$; we call any such good $j$ \emph{expensive} (or \emph{weakly oversupplied}) at price vector $\mathbf{p}$.

For each integer $k \ge 1$, let $\mathcal{T}_k$ be a triangulation of $\Delta^{m-1}$ of mesh size $\max_{\sigma \in \mathcal{T}_k} \operatorname{diam}(\sigma) \le 1/k$. Label each vertex $\mathbf{p} \in V(\mathcal{T}_k)$ with the smallest index $\ell(\mathbf{p}) = j \in \{1, \dots, m\}$ such that $p_j > 0$ and $z_j(\mathbf{p}) \le 1$. By construction, if $\mathbf{p}$ lies on facet $F_j$ ($p_j = 0$), then $\ell(\mathbf{p}) \neq j$ because we only assign labels from $\operatorname{supp}(\mathbf{p}) = \{r : p_r > 0\}$ (and economically, a free good with $p_j = 0$ has $z_j(\mathbf{p}) > 1$ and cannot be expensive). At each principal vertex $\mathbf{e}_j$, $\operatorname{supp}(\mathbf{e}_j) = \{j\}$, so $\ell(\mathbf{e}_j) = j$. Thus $\ell$ is a valid Sperner labeling of $\mathcal{T}_k$.

By Lemma~\ref{lem:sperner}, the triangulation $\mathcal{T}_k$ contains a panchromatic elementary simplex $\sigma_k = \operatorname{conv}\{\mathbf{p}^{(k,1)}, \dots, \mathbf{p}^{(k,m)}\}$ with $\ell(\mathbf{p}^{(k,j)}) = j$ for each $j \in \{1,\dots,m\}$. By compactness of $\Delta^{m-1}$, passing to a convergent subsequence as $k \to \infty$, all $m$ vertices $\mathbf{p}^{(k,1)}, \dots, \mathbf{p}^{(k,m)}$ converge to a common limit point $\mathbf{p}^* \in \Delta^{m-1}$. By continuity of $z_j$ and the closedness of the condition $z_j(\mathbf{p}^{(k,j)}) \le 1$, we obtain $z_j(\mathbf{p}^*) \le 1$ for every $j \in \{1, \dots, m\}$ (and $p_j^* > 0$ for all $j$ by boundary desirability). Combining $z_j(\mathbf{p}^*) \le 1$ for all $j$ with $\sum_{j=1}^m p_j^* (1 - z_j(\mathbf{p}^*)) = 0$ and $p_j^* > 0$ forces $z_j(\mathbf{p}^*) = 1$ for every $j \in \mathcal{M}$. Hence $\mathbf{p}^*$ is a market-clearing price vector.
\end{proof}

While Proposition~\ref{prop:sperner-ce} guarantees existence, finding a panchromatic simplex via Sperner subdivision is $\PPAD$-complete in general dimensions. For linear Fisher markets, however, the convex structure of Theorem~\ref{thm:eg-kkt} enables polynomial-time combinatorial algorithms.

\subsection{The DPSV Primal-Dual Algorithm and Orlin's Strongly Polynomial Algorithm}
\label{subsec:dpsv}

Devanur, Papadimitriou, Saberi, and Vazirani (DPSV) \cite{Devanur2008,Vazirani2007} developed the first combinatorial polynomial-time algorithm for linear Fisher markets by interpreting the KKT conditions of the Eisenberg--Gale program~\eqref{eq:eg-program} through a parametric maximum-flow network.

\paragraph{The equality subgraph flow network $G(\mathbf{p})$.}
Fix a strictly positive price vector $\mathbf{p} = (p_1, \dots, p_m) > \mathbf{0}$. Recall from~\eqref{eq:bang-per-buck} that buyer $i$'s maximum bang-per-buck at prices $\mathbf{p}$ is $\alpha_i(\mathbf{p}) = \max_{k \in \mathcal{M}} (v_{ik}/p_k)$. Construct the directed bipartite flow network $G(\mathbf{p}) = (\{s, t\} \cup \mathcal{M} \cup \mathcal{N}, E(\mathbf{p}))$ as follows (illustrated in Figure~\ref{fig:flow-network}):
\begin{enumerate}[label=(\roman*),topsep=2pt,itemsep=2pt]
  \item Add a directed edge $(s, j)$ from the source $s$ to each good $j \in \mathcal{M}$ with capacity $c(s, j) = p_j$ (the total revenue good $j$ must generate when its supply is $C_j = 1$).
  \item Add a directed edge $(j, i)$ from good $j \in \mathcal{M}$ to buyer $i \in \mathcal{N}$ with capacity $c(j, i) = \infty$ if and only if good $j$ maximizes buyer $i$'s bang-per-buck at prices $\mathbf{p}$, i.e.,
  \[
    (j, i) \in E(\mathbf{p}) \iff \frac{v_{ij}}{p_j} = \alpha_i(\mathbf{p}) = \max_{k \in \mathcal{M}} \frac{v_{ik}}{p_k}.
  \]
  These edges form the \emph{equality subgraph}.
  \item Add a directed edge $(i, t)$ from each buyer $i \in \mathcal{N}$ to the sink $t$ with capacity $c(i, t) = B_i$ (the total money possessed by buyer $i$).
\end{enumerate}

\begin{figure}[t]
\centering
\begin{tikzpicture}[
  node distance=12mm,
  vertex/.style={circle, draw=black!85, line width=0.6pt, fill=white, minimum size=7.5mm, inner sep=1pt, font=\small},
  arr/.style={-{Stealth[length=2.2mm,width=1.5mm]}, line width=0.6pt, draw=black!85},
  lbl/.style={font=\footnotesize, fill=white, inner sep=1.2pt}
]
  \node[vertex] (s) at (0, -2.25) {$s$};
  \node[vertex] (t) at (9.2, -2.25) {$t$};

  \node[vertex] (m1) at (2.8,  0.0) {$g_1$};
  \node[vertex] (m2) at (2.8, -1.5) {$g_2$};
  \node[vertex] (m3) at (2.8, -3.0) {$g_3$};
  \node[vertex] (m4) at (2.8, -4.5) {$g_4$};

  \node[vertex] (b1) at (6.4, -0.5) {$b_1$};
  \node[vertex] (b2) at (6.4, -2.25) {$b_2$};
  \node[vertex] (b3) at (6.4, -4.0) {$b_3$};

  \draw[arr] (s) -- node[lbl, pos=0.55] {$p_1$} (m1);
  \draw[arr] (s) -- node[lbl, pos=0.55] {$p_2$} (m2);
  \draw[arr] (s) -- node[lbl, pos=0.55] {$p_3$} (m3);
  \draw[arr] (s) -- node[lbl, pos=0.55] {$p_4$} (m4);

  \draw[arr] (m1) -- node[lbl, pos=0.45] {$\infty$} (b1);
  \draw[arr] (m2) -- node[lbl, pos=0.45] {$\infty$} (b1);
  \draw[arr] (m2) -- node[lbl, pos=0.50] {$\infty$} (b2);
  \draw[arr] (m3) -- node[lbl, pos=0.45] {$\infty$} (b2);
  \draw[arr] (m3) -- node[lbl, pos=0.50] {$\infty$} (b3);
  \draw[arr] (m4) -- node[lbl, pos=0.45] {$\infty$} (b3);

  \draw[arr] (b1) -- node[lbl, pos=0.45] {$B_1$} (t);
  \draw[arr] (b2) -- node[lbl, pos=0.45] {$B_2$} (t);
  \draw[arr] (b3) -- node[lbl, pos=0.45] {$B_3$} (t);

  \node[font=\footnotesize\bfseries] at (2.8, 0.75) {Goods $\mathcal{M}$};
  \node[font=\footnotesize\bfseries] at (6.4, 0.25) {Buyers $\mathcal{N}$};
\end{tikzpicture}
\caption{The DPSV equality subgraph network $G(\mathbf{p})$ for $m = 4$ goods $\mathcal{M} = \{g_1, g_2, g_3, g_4\}$ and $n = 3$ buyers $\mathcal{N} = \{b_1, b_2, b_3\}$. Edges from $s$ to $g_j$ carry capacity $p_j$; edges $(g_j, b_i)$ exist with capacity $\infty$ whenever $v_{ij}/p_j = \alpha_i(\mathbf{p})$; and edges from $b_i$ to $t$ carry capacity $B_i$.}
\label{fig:flow-network}
\end{figure}

In $G(\mathbf{p})$, a flow $f: E(\mathbf{p}) \to \mathbb{R}_{\ge 0}$ represents money flowing from goods to buyers: $f(j, i)$ is the dollar amount buyer $i$ spends on good $j$, corresponding to physical allocation $x_{ij} = f(j, i)/p_j$.

\begin{lemma}[Min-Cut Characterization of Competitive Equilibrium \cite{Devanur2008,Vazirani2007}]
\label{lem:mincut-ce}
A price vector $\mathbf{p} > \mathbf{0}$ is an equilibrium price vector of the linear Fisher market if and only if the two cuts
\[
  (\{s\},\; \mathcal{M} \cup \mathcal{N} \cup \{t\})
  \qquad\text{and}\qquad
  (\{s\} \cup \mathcal{M} \cup \mathcal{N},\; \{t\})
\]
are both minimum $s$--$t$ cuts in $G(\mathbf{p})$. When this holds, any maximum $s$--$t$ flow $f^*$ saturates all source edges ($f^*(s, j) = p_j$) and all sink edges ($f^*(i, t) = B_i$), and setting $x_{ij}^* = f^*(j, i)/p_j$ yields an equilibrium allocation.
\end{lemma}
\begin{proof}
The capacity of $(\{s\}, \mathcal{M} \cup \mathcal{N} \cup \{t\})$ is $\sum_{j \in \mathcal{M}} p_j$, and the capacity of $(\{s\} \cup \mathcal{M} \cup \mathcal{N}, \{t\})$ is $\sum_{i \in \mathcal{N}} B_i$. By the max-flow min-cut theorem, both cuts are minimum $s$--$t$ cuts simultaneously if and only if the maximum $s$--$t$ flow value equals $\sum_{j \in \mathcal{M}} p_j = \sum_{i \in \mathcal{N}} B_i$. In any such maximum flow $f^*$, every source edge $(s, j)$ is saturated ($\sum_{i \in \mathcal{N}} f^*(j, i) = p_j$, so $\sum_{i \in \mathcal{N}} x_{ij}^* = 1$), every sink edge $(i, t)$ is saturated ($\sum_{j \in \mathcal{M}} p_j x_{ij}^* = B_i$), and flow is sent only along edges $(j, i) \in E(\mathbf{p})$ where $v_{ij}/p_j = \alpha_i(\mathbf{p})$. These are precisely the KKT conditions (i)--(iv) of Theorem~\ref{thm:eg-kkt}.
\end{proof}

\begin{algorithm}[t]
\caption{DPSV Primal-Dual Ascending-Price Algorithm Sketch \cite{Devanur2008,Vazirani2007}}
\label{alg:dpsv}
\begin{algorithmic}[1]
\Require Buyers $\mathcal{N} = \{1,\dots,n\}$ with integer budgets $B_i > 0$, goods $\mathcal{M} = \{1,\dots,m\}$ with unit supplies and integer valuations $v_{ij} \ge 0$.
\Ensure Equilibrium price vector $\mathbf{p}^*$ and allocation $x_{ij}^* = f^*(j, i)/p_j^*$.
\State Initialize low prices $p_j \leftarrow \frac{1}{m} \min_{i \in \mathcal{N}} B_i$ so that $(\{s\}, \mathcal{M} \cup \mathcal{N} \cup \{t\})$ is the unique min-cut in $G(\mathbf{p})$.
\While{$\sum_{j \in \mathcal{M}} p_j < \sum_{i \in \mathcal{N}} B_i$}
  \State Construct the equality subgraph $G(\mathbf{p})$ and compute a \emph{balanced} maximum flow $f$ that minimizes the $\ell_2$-norm of buyers' surplus money $\gamma_i = B_i - f(i, t)$.
  \State Identify the active subgraph $\mathcal{A} \subseteq \mathcal{M} \cup \mathcal{N}$ of goods and buyers with positive surplus and scale prices of goods $j \in \mathcal{A} \cap \mathcal{M}$ multiplicatively by $p_j \leftarrow x \cdot p_j$ for $x \ge 1$.
  \State Increase $x$ continuously from $1$ until either:
  \Statex \hspace{1.8em}\textbf{Event 1 (New equality edge):} a buyer $i \in \mathcal{A} \cap \mathcal{N}$ becomes equally happy with a good $j \notin \mathcal{A} \cap \mathcal{M}$, adding $(j, i)$ to $E(\mathbf{p})$; or
  \Statex \hspace{1.8em}\textbf{Event 2 (New tight set):} a subset $S \subseteq \mathcal{A} \cap \mathcal{M}$ of goods becomes \emph{tight}, meaning $\sum_{j \in S} x \cdot p_j = \sum_{i \in \Gamma(S)} B_i$, where $\Gamma(S) \subseteq \mathcal{A} \cap \mathcal{N}$ are the buyers with edges only into $S$. Freeze prices on $S$ and remove $S \cup \Gamma(S)$ from the active set.
\EndWhile
\State \Return $\mathbf{p}^* = \mathbf{p}$ and $x_{ij}^* = f^*(j, i)/p_j^*$.
\end{algorithmic}
\end{algorithm}

\noindent\textbf{Primal-dual price scaling and strongly polynomial complexity.}
Algorithm~\ref{alg:dpsv} summarizes the DPSV primal-dual method \cite{Devanur2008,Vazirani2007}. Starting from low prices $p_j = \frac{1}{m}\min_i B_i$ at which the source cut $(\{s\}, \mathcal{M} \cup \mathcal{N} \cup \{t\})$ is a minimum cut (every good is fully sold, while buyers retain unspent surplus $\gamma_i = B_i - f(i, t) \ge 0$), the algorithm maintains source-side saturation and raises prices of active goods multiplicatively ($p_j \leftarrow x \cdot p_j$) to preserve existing equality edges inside the active set until all buyer surpluses vanish. Using a potential function based on the $\ell_2$-norm of the surplus vector in a balanced flow, Devanur et al.\ \cite{Devanur2008} proved that Algorithm~\ref{alg:dpsv} terminates with the exact rational equilibrium in weakly polynomial time
\[
  O\bigl((n + m)^8 \log v_{\max} + (n + m)^7 \log B_{\max}\bigr),
\]
where $v_{\max} = \max_{i,j} v_{ij}$ and $B_{\max} = \max_i B_i$.

Subsequently, Orlin \cite{Orlin2010} resolved the longstanding open problem of whether linear Fisher market equilibria can be computed in \emph{strongly polynomial time} (a number of arithmetic operations polynomial in $n$ and $m$ alone, independent of the bit-lengths of $v_{ij}$ and $B_i$). By combining a scaling refinement of the Eisenberg--Gale dual with abundant-good identification that permanently fixes variables across $O(n + m)$ scaling phases, Orlin \cite{Orlin2010} achieved a strongly polynomial running time of
\[
  O\bigl((n + m)^4 \log(n + m)\bigr).
\]
In parallel, Fleischer, Garg, Kapoor, Khandekar, and Saberi \cite{Fleischer2008} developed a fast distributed multiplicative-weights auction algorithm that computes an $\varepsilon$-approximate market equilibrium in $O(\frac{1}{\varepsilon^2}(n+m)\log(n+m))$ phases without requiring a central coordinator to solve max-flow instances, and Devanur and Kannan \cite{DevanurKannan2008} gave polynomial-time cell-decomposition algorithms for Arrow--Debreu and piecewise-linear concave (PLC) markets when the number of goods or agents is fixed.

\section{Indivisible Goods and Competitive Equilibrium from Equal Incomes}
\label{sec:indivisible}

The existence and polynomial-time tractability of competitive equilibrium in Section~\ref{sec:divisible} rely crucially on the divisibility of goods ($x_{ij} \in \mathbb{R}_{\ge 0}$). When goods are indivisible ($x_{ij} \in \mathbb{Z}_{\ge 0}$ and supplies $C_j \in \mathbb{Z}_{>0}$ are integral), both existence and computational tractability change fundamentally.

\subsection{Indivisible Fisher Markets and \texorpdfstring{$\varepsilon$}{epsilon}-Approximate Equilibria}

In a discrete Fisher market, an allocation $\mathbf{x} = (x_{ij}) \in \mathbb{Z}_{\ge 0}^{n \times m}$ must assign an integer number of units of each good $j \in \mathcal{M}$ to each buyer $i \in \mathcal{N}$. Even in the simplest case of $m = 1$ indivisible item ($C_1 = 1$) and $n = 2$ buyers with positive budgets $B_1 = B_2 = 1$ and $v_{11} = v_{21} = 1$, no exact competitive equilibrium exists: if $p_1 \le 1$, both buyers demand the item (total demand $2 > 1$), whereas if $p_1 > 1$, neither buyer can afford it (total demand $0 < 1$).

More generally, deciding whether an instance happens to admit an exact integral competitive equilibrium is computationally intractable.

\begin{theorem}[Deng, Papadimitriou, and Safra \cite{DengPapadimitriouSafra2003}]
\label{thm:dps03}
In a discrete Fisher market with indivisible goods and linear utilities:
\begin{enumerate}[label=(\alph*),topsep=2pt,itemsep=2pt]
  \item Deciding whether an exact competitive equilibrium exists is $\NP$-hard, even when there are only $n = 3$ buyers.
  \item For any fixed number of goods $m$ and error tolerance $\varepsilon \in (0, 1)$, there is an algorithm running in time
  \[
    O\!\left(\left(\frac{n^2 \log(\max_{j \in \mathcal{M}} C_j)}{\varepsilon}\right)^{\!2m}\right)
  \]
  (polynomial in the input size and $1/\varepsilon$ for fixed $m$) that computes an \emph{$\varepsilon$-approximate competitive equilibrium}, in which each buyer receives an $(1-\varepsilon)$-optimal bundle within her budget and the aggregate allocation satisfies
  \begin{equation}
    (1 - \varepsilon) C_j \;\le\; \sum_{i \in \mathcal{N}} x_{ij} \;\le\; C_j, \qquad \forall j \in \mathcal{M}.
    \label{eq:eps-approx-ce}
  \end{equation}
\end{enumerate}
\end{theorem}

The algorithm of Theorem~\ref{thm:dps03}(b) discretizes the $m$-dimensional price space into a logarithmic grid of size $O((\frac{n \log(\max_j C_j)}{\varepsilon})^m)$ and uses dynamic programming across the $n$ buyers to test whether rounded integer demands can pack into~\eqref{eq:eps-approx-ce}.

\subsection{Competitive Equilibrium from Equal Incomes (CEEI)}
\label{subsec:ceei}

When indivisible goods must be divided fairly among agents without monetary transfers, assigning each agent an equal artificial budget $B_1 = B_2 = \dots = B_n = 1$ yields the \emph{Competitive Equilibrium from Equal Incomes (CEEI)} mechanism \cite{Budish2011,Aziz2015,BouveretLemaitre2016}. An allocation generated by a CEEI is simultaneously Pareto efficient and \emph{envy-free} (since every agent faces the same price vector $\mathbf{p}$ and has the same budget $1$, every agent's chosen bundle is at least as preferred as any other agent's bundle).

\paragraph{Complexity of exact CEEI and \prob{CEEI-FRAC}.}
Bouveret and Lema\^\i tre \cite{BouveretLemaitre2016} positioned CEEI at the top of a fairness hierarchy for additive valuations:
\[
  \text{CEEI} \implies \text{Envy-Freeness} \implies \text{Proportionality} \implies \text{Max-Min Share},
\]
and posed the complexity of testing whether an additive indivisible instance admits a CEEI. Aziz \cite{Aziz2015} and Br\^anzei, Hosseini, and Miltersen \cite{Branzei2015} resolved this question completely:

\begin{proposition}[Complexity of CEEI with Additive Valuations \cite{Aziz2015,Branzei2015}]
\label{prop:aziz-ceei}
Consider a discrete allocation instance with $n$ agents having equal budgets $B_i = 1$, $m$ indivisible unit-supply goods, and additive valuations $u_i(\mathbf{x}_i) = \sum_{j=1}^m v_{ij} x_{ij}$:
\begin{enumerate}[label=(\arabic*),topsep=2pt,itemsep=2pt]
  \item \textbf{Two-agent weak $\NP$-hardness:} For $n = 2$ agents and $m$ items with identical integer valuations $v_{1j} = v_{2j} = a_j \in \mathbb{Z}_{>0}$, a CEEI exists if and only if $\{a_1, \dots, a_m\}$ can be partitioned into two subsets of equal sum $\frac{1}{2}\sum_{j=1}^m a_j$ (by setting $p_j = a_j / (\frac{1}{2}\sum_k a_k)$). Thus deciding CEEI existence for $n = 2$ agents is weakly $\NP$-hard by reduction from \prob{Partition} \cite{Aziz2015}.
  \item \textbf{General strong $\NP$-hardness:} For $n$ agents and $m = 3n$ items, deciding whether a CEEI exists is strongly $\NP$-hard by reduction from \prob{3-Partition} \cite{Aziz2015}. Furthermore, even verifying whether a given integral allocation $\mathbf{x}^* \in \{0,1\}^{n \times m}$ can be supported by some equilibrium price vector $\mathbf{p}^*$ under ordinary indivisible bundle demand is $\coNP$-hard \cite{Branzei2015}, and Heinen, Nguyen, and Rothe \cite{Heinen2015} extended these hardness results to $k$-additive utility functions.
  \item \textbf{Tractability of \prob{CEEI-FRAC}:} If the demand condition is strengthened to \prob{CEEI-FRAC} \cite{Aziz2015}---which requires that the integral allocation $\mathbf{x}_i^* \in \{0,1\}^m$ also be optimal for each buyer $i$ among all \emph{fractional} bundles $\mathbf{y}_i \in [0, 1]^m$ satisfying $\sum_{j=1}^m p_j y_{ij} \le 1$, i.e.,
  \[
    \mathbf{x}_i^* \in \arg\max_{\mathbf{y}_i \in [0,1]^m} \Biggl\{\sum_{j \in \mathcal{M}} v_{ij} y_{ij} \;\Big|\; \sum_{j \in \mathcal{M}} p_j y_{ij} \le 1\Biggr\}
  \]
  ---then verifying whether a given allocation $\mathbf{x}^*$ is supported by \prob{CEEI-FRAC} prices is solvable in polynomial time via linear programming (equivalent to the Eisenberg--Gale KKT conditions of Theorem~\ref{thm:eg-kkt}), placing \prob{CEEI-FRAC} in $\NP$.
\end{enumerate}
\end{proposition}

\begin{table}[t]
\centering
\caption{Summary of equilibrium existence and computational complexity across divisible and indivisible Fisher market models.}
\label{tab:market-complexity}
\footnotesize
\setlength{\tabcolsep}{1.8pt}
\begin{tabular}{@{}lllll@{}}
\toprule
Market Model & Goods & Utility / Regime & Existence & Complexity \\
\midrule
Linear Fisher & Divisible & Linear ($\sum_j v_{ij} x_{ij}$) & Always \cite{EisenbergGale1959} & $O((n+m)^4 \log(n+m))$ \cite{Orlin2010} \\
EG / CES ($\rho < 1$) & Divisible & Homogeneous & Always \cite{Eisenberg1961} & Convex program / poly-time \cite{JainVazirani2010} \\
Leontief Fisher & Divisible & $\min_j (x_{ij}/v_{ij})$ & Always & Convex program \cite{CodenottiVaradarajan2004} \\
Discrete Fisher & Indivisible & Linear ($n \ge 3$) & Conditional & $\NP$-hard ($\varepsilon$-CE: poly \cite{DengPapadimitriouSafra2003}) \\
Exact CEEI ($n = 2$) & Indivisible & Additive & Conditional & Weakly $\NP$-hard (\prob{Partition}) \cite{Aziz2015} \\
Exact CEEI ($n$ agents) & Indivisible & Additive & Conditional & Strongly $\NP$-hard (\prob{3-Partition}) \cite{Aziz2015} \\
\prob{CEEI-FRAC} & Indivisible & Fractional demand & Conditional & Verify in $\Pclass$; decide in $\NP$ \cite{Aziz2015} \\
Budish A-CEEI & Indivisible & Ordinal bundles & Always \cite{Budish2011} & $\PPAD$-complete \cite{OthmanPapadimitriouRubinstein2016} \\
\bottomrule
\end{tabular}
\end{table}

\paragraph{Budish's Approximate CEEI (A-CEEI), Maximum Nash Welfare, and Generic Budgets.}
Because exact CEEI fails to exist for indivisible goods with equal budgets, three major relaxations have been developed in the market design literature (summarized in Table~\ref{tab:market-complexity}):
\begin{enumerate}[label=(\arabic*),topsep=2pt,itemsep=2pt]
  \item \textbf{Budish's Approximate CEEI (A-CEEI) \cite{Budish2011}:}
  Motivated by combinatorial course allocation (such as the Wharton School course bidding auction, where standard bidding mechanisms incentivize strategic manipulation), Budish \cite{Budish2011} proved that for arbitrary ordinal preferences over bundles of at most $k$ items, if budgets are perturbed within $[1, 1 + \beta]$ for arbitrarily small $\beta > 0$, there always exists an \emph{Approximate CEEI} whose market clearing error is bounded in $\ell_2$-norm by $\frac{\sqrt{k m}}{2}$ independent of the number of agents $n$, and which is envy-free up to one good (EF1) and strategyproof in the large. However, Othman, Papadimitriou, and Rubinstein \cite{OthmanPapadimitriouRubinstein2016} proved that computing an A-CEEI is $\PPAD$-complete.
  \item \textbf{Maximum Nash Welfare and Integral Proximity \cite{Caragiannis2016,BarmanKrishnamurthy2019}:}
  Caragiannis et al.\ \cite{Caragiannis2016} showed that maximizing the integral Nash social welfare $\prod_{i=1}^n u_i(\mathbf{x}_i)$ directly over integral allocations $\mathbf{x} \in \{0,1\}^{n \times m}$ serves as the discrete analogue of the Eisenberg--Gale program, always producing an allocation that is Pareto efficient and envy-free up to one good (EF1). Complementing this, Barman and Krishnamurthy \cite{BarmanKrishnamurthy2019} proved that any fractional Fisher market equilibrium with additive utilities can be rounded to an exact integral competitive equilibrium of a nearby Fisher market whose budgets $B_i'$ differ from $B_i$ by at most the maximum price $\max_j p_j^*$ of a single item.
  \item \textbf{Generic Budgets \cite{BabaioffNisanTalgamCohen2021}:}
  Babaioff, Nisan, and Talgam-Cohen \cite{BabaioffNisanTalgamCohen2021} showed that the non-existence of CEEI is partly an artifact of the knife-edge symmetry $B_1 = \dots = B_n$: when budgets $(B_1, \dots, B_n)$ are in general position (outside a measure-zero union of hyperplanes), exact competitive equilibria exist for all markets with $m = 3$ indivisible goods and for $m = 4$ goods with $n = 2$ buyers, though non-existence can persist on open budget sets for $m \ge 5$ goods.
\end{enumerate}

\section{Fisher Markets with Positional (Snob-Effect) Externalities}
\label{sec:snob}

In the classical Fisher and Eisenberg--Gale models of Sections~\ref{sec:preliminaries}--\ref{sec:indivisible}, buyer $i$'s utility $u_i(\mathbf{x}_i)$ depends only on her own consumption bundle $\mathbf{x}_i = (x_{i1}, \dots, x_{im})$, independent of how much of good $j$ is consumed by other buyers. In a foundational microeconomic study, Leibenstein \cite{Leibenstein1950} distinguished functional demand from \emph{non-functional positional demand}, identifying the \emph{snob effect} as the phenomenon in which a consumer's valuation for a good increases with its exclusivity and decreases with the aggregate share of the good held by others (as in luxury goods, limited-edition collectibles, status credentials, or congestible club goods).

In this section, we formulate a divisible Fisher market model that incorporates both intrinsic buyer preferences $v_{ij} \ge 0$ and buyer-specific snob sensitivity $\alpha_i \ge 0$, and we establish two rigorous structural results:
\begin{enumerate}[label=(\arabic*),topsep=2pt,itemsep=2pt]
  \item We prove that for any $\alpha_i > 0$, each positional utility term is a \emph{convex} function of the allocation vector on $\mathbb{R}_{\ge 0}^n$, so utilitarian social welfare is maximized at exclusive corner allocations where each good is allocated in its entirety to a single buyer (Theorem~\ref{thm:snob-welfare}).
  \item We analyze decentralized market clearing under positional externalities, proving that in a single-good snob market, budget-proportional allocation is the unique price-taking competitive equilibrium, and that decentralized competition suffers a Price of Anarchy of $\PoA = n^\alpha$ relative to the social optimum among $n$ symmetric buyers (Theorem~\ref{thm:snob-equilibrium}).
\end{enumerate}

\subsection{The Positional Fisher Market Model}

Let $\mathcal{N} = \{1, \dots, n\}$ be a set of $n \ge 2$ buyers with budgets $B_i > 0$, and let $\mathcal{M} = \{1, \dots, m\}$ be a set of $m \ge 1$ divisible goods with supplies $C_j > 0$. For an allocation $\mathbf{x} = (x_{ij}) \in \mathbb{R}_{\ge 0}^{n \times m}$, let
\[
  X_j := \sum_{k \in \mathcal{N}} x_{kj}
\]
denote the aggregate consumption of good $j \in \mathcal{M}$, and define buyer $i$'s \emph{consumption share} of good $j$ by
\[
  s_{ij}(\mathbf{x}) := \begin{cases}
    \dfrac{x_{ij}}{X_j} & \text{if } X_j > 0, \\[0.6em]
    0                   & \text{if } X_j = 0.
  \end{cases}
\]
A purely positional utility function $\sum_{j \in \mathcal{M}} s_{ij}(\mathbf{x})$ captures exclusivity ($s_{ij} = 1$ when buyer $i$ is the sole owner of good $j$, and $s_{ij} = 1/n$ under equal sharing), but ignores buyer $i$'s intrinsic baseline valuation $v_{ij} \ge 0$ per unit of good $j$ and the absolute quantity $x_{ij}$ consumed. To combine intrinsic utility with positional exclusivity, we associate each buyer $i \in \mathcal{N}$ with a \emph{snob sensitivity parameter} $\alpha_i \ge 0$ and define her positional utility function:

\begin{definition}[Positional Snob-Effect Utility]
\label{def:snob-utility}
Given intrinsic valuations $v_{ij} \ge 0$ and sensitivity parameter $\alpha_i \ge 0$, the utility of buyer $i \in \mathcal{N}$ at allocation $\mathbf{x} = (x_{ij}) \in \mathbb{R}_{\ge 0}^{n \times m}$ is
\begin{equation}
  U_i(\mathbf{x}) = \sum_{j \in \mathcal{M}} v_{ij} x_{ij} \Biggl(\frac{x_{ij}}{\sum_{k \in \mathcal{N}} x_{kj}}\Biggr)^{\!\alpha_i}
  = \sum_{j \in \mathcal{M}} v_{ij} x_{ij} \, s_{ij}(\mathbf{x})^{\alpha_i},
  \label{eq:snob-utility}
\end{equation}
where the $j$-th summand is defined to be $0$ whenever $x_{ij} = 0$.
\end{definition}

Observe three immediate properties of Definition~\ref{def:snob-utility}:
\begin{enumerate}[label=(\roman*),topsep=2pt,itemsep=2pt]
  \item \textbf{Reduction to the linear Fisher market at $\alpha_i = 0$:} When $\alpha_i = 0$ for all $i \in \mathcal{N}$, $(x_{ij}/X_j)^0 = 1$ and $U_i(\mathbf{x}) = \sum_{j \in \mathcal{M}} v_{ij} x_{ij}$, recovering the classical linear Fisher market of Section~\ref{sec:preliminaries}.
  \item \textbf{Monotonicity in own consumption:} For fixed opponent allocations $X_{-i,j} = \sum_{k \neq i} x_{kj} \ge 0$ and $\alpha_i \ge 0$, both $x_{ij}$ and $s_{ij} = \frac{x_{ij}}{x_{ij} + X_{-i,j}}$ are strictly increasing in $x_{ij} > 0$, with partial derivative
  \begin{equation}
    \frac{\partial U_i}{\partial x_{ij}}(\mathbf{x})
    = v_{ij} s_{ij}^{\alpha_i} \bigl(1 + \alpha_i(1 - s_{ij})\bigr) > 0
    \qquad (\text{for } x_{ij} > 0,\; v_{ij} > 0).
    \label{eq:snob-marginal}
  \end{equation}
  \item \textbf{Negative consumption externality:} For any $k \neq i$ and $\alpha_i > 0$,
  \[
    \frac{\partial U_i}{\partial x_{kj}}(\mathbf{x})
    = -\alpha_i v_{ij} s_{ij}^{\alpha_i + 1} < 0
    \qquad (\text{for } x_{ij} > 0,\; v_{ij} > 0),
  \]
  capturing the snob externality: an increase in another buyer's holdings of good $j$ strictly reduces buyer $i$'s utility from good $j$.
\end{enumerate}

\subsection{Convexity and Exclusive Social Welfare Optimality}

Given prices $\mathbf{p} = (p_1, \dots, p_m) \in \mathbb{R}_{\ge 0}^m$, budgets $B_i > 0$, and supplies $C_j > 0$, consider the utilitarian social welfare maximization problem over the feasible allocation polytope $\mathcal{F}(\mathbf{p})$:
\begin{equation}
\begin{aligned}
  &\text{maximize}_{\mathbf{x} \in \mathbb{R}^{n \times m}} &&\SW(\mathbf{x}) = \sum_{i \in \mathcal{N}} \sum_{j \in \mathcal{M}} v_{ij} x_{ij} \Biggl(\frac{x_{ij}}{\sum_{k \in \mathcal{N}} x_{kj}}\Biggr)^{\!\alpha_i} \\
  &\text{subject to} &&\sum_{j \in \mathcal{M}} p_j x_{ij} \le B_i, \quad \forall i \in \mathcal{N}, \\
  &                  &&\sum_{i \in \mathcal{N}} x_{ij} \le C_j,     \quad \forall j \in \mathcal{M}, \qquad x_{ij} \ge 0 \;\;\forall i \in \mathcal{N},\; \forall j \in \mathcal{M}.
\end{aligned}
\label{eq:snob-welfare-prog}
\end{equation}

We first prove that for every $\alpha_i > 0$, the positional utility function $U_i(\mathbf{x})$ is \emph{convex} (rather than concave) on $\mathbb{R}_{\ge 0}^{n \times m}$, and therefore utilitarian social welfare $\SW(\mathbf{x})$ always achieves its global maximum at an extreme point of the feasible polytope.

\begin{theorem}[Convexity of Positional Utility and Exclusive Welfare Optimality]
\label{thm:snob-welfare}
Let $\alpha_i > 0$ and $v_{ij} \ge 0$ for all $i \in \mathcal{N}$ and $j \in \mathcal{M}$. Then:
\begin{enumerate}[label=(\alph*),topsep=2pt,itemsep=2pt]
  \item For each buyer $i \in \mathcal{N}$ and good $j \in \mathcal{M}$, the function $g_{ij}: \mathbb{R}_{\ge 0}^n \to \mathbb{R}_{\ge 0}$ defined by
  \[
    g_{ij}(x_{1j}, \dots, x_{nj}) = v_{ij} x_{ij}^{1 + \alpha_i} \Biggl(\sum_{k=1}^n x_{kj}\Biggr)^{\!-\alpha_i}
  \]
  (with $g_{ij}(\mathbf{0}) = 0$) is continuous and \emph{convex} on $\mathbb{R}_{\ge 0}^n$, and when $v_{ij} > 0$ it is strictly convex along any line segment in $\mathbb{R}_{>0}^n$ on which the consumption share $s_{ij} = x_{ij} / \sum_{k=1}^n x_{kj}$ is non-constant.
  \item Consequently, the social welfare objective $\SW(\mathbf{x}) = \sum_{i \in \mathcal{N}} \sum_{j \in \mathcal{M}} g_{ij}(\mathbf{x}_{\cdot, j})$ is convex on the compact polytope $\mathcal{F}(\mathbf{p})$ of~\eqref{eq:snob-welfare-prog} and attains its global maximum at an extreme point (vertex) of $\mathcal{F}(\mathbf{p})$.
  \item In particular, if budgets are non-binding (or in the unpriced supply-constrained problem $\sum_{i \in \mathcal{N}} x_{ij} \le C_j$, $x_{ij} \ge 0$), an optimal allocation assigns the entire supply $C_j$ of each good $j \in \mathcal{M}$ \emph{exclusively to a single buyer} $i^*(j) \in \arg\max_{i \in \mathcal{N}} v_{ij}$, achieving maximum social welfare
  \begin{equation}
    \max_{\substack{\mathbf{x} \ge \mathbf{0} \\ \sum_i x_{ij} \le C_j}} \SW(\mathbf{x})
    = \sum_{j \in \mathcal{M}} C_j \max_{i \in \mathcal{N}} v_{ij}.
    \label{eq:snob-opt-val}
  \end{equation}
\end{enumerate}
\end{theorem}
\begin{proof}
Fix $i \in \mathcal{N}$ and $j \in \mathcal{M}$ with $v_{ij} > 0$ and $\alpha_i > 0$. Write $x = x_{ij} \ge 0$ and $S = \sum_{k=1}^n x_{kj} \ge x$ for the aggregate allocation of good $j$. Since $S$ is a linear function of $(x_{1j}, \dots, x_{nj})$, to prove that $g_{ij}$ is convex on $\mathbb{R}_{\ge 0}^n$ it suffices to prove that the bivariate function
\[
  \phi(x, S) := x^{1 + \alpha_i} S^{-\alpha_i}
\]
is convex on the cone $\mathcal{K} = \{(x, S) \in \mathbb{R}_{>0}^2 : x \le S\}$ (continuity up to the boundary follows from $0 \le \phi(x, S) \le x$). Computing the first and second partial derivatives of $\phi(x, S)$ on $\mathcal{K}$ gives
\begin{align*}
  \frac{\partial \phi}{\partial x} &= (1 + \alpha_i) \left(\frac{x}{S}\right)^{\!\alpha_i},
  & \frac{\partial \phi}{\partial S} &= -\alpha_i \left(\frac{x}{S}\right)^{\!\alpha_i + 1}, \\[0.4em]
  \frac{\partial^2 \phi}{\partial x^2} &= \alpha_i(1 + \alpha_i) x^{\alpha_i - 1} S^{-\alpha_i},
  & \frac{\partial^2 \phi}{\partial x \partial S} &= -\alpha_i(1 + \alpha_i) x^{\alpha_i} S^{-\alpha_i - 1}, \\
  \frac{\partial^2 \phi}{\partial S^2} &= \alpha_i(1 + \alpha_i) x^{\alpha_i + 1} S^{-\alpha_i - 2}.
\end{align*}
Thus, for any displacement $(\delta x, \delta S) \in \mathbb{R}^2$, the quadratic form of the Hessian $\nabla^2 \phi(x, S)$ is
\begin{equation}
  (\delta x, \delta S) \nabla^2 \phi(x, S) \begin{pmatrix} \delta x \\ \delta S \end{pmatrix}
  = \alpha_i(1 + \alpha_i) x^{\alpha_i - 1} S^{-\alpha_i - 2} \bigl(S\,\delta x - x\,\delta S\bigr)^2 \;\ge\; 0,
  \label{eq:snob-hessian}
\end{equation}
with strict positivity whenever $S\,\delta x \neq x\,\delta S$, i.e., whenever $\delta(x/S) = \frac{S\,\delta x - x\,\delta S}{S^2} \neq 0$. This proves part~(a).

Since $\SW(\mathbf{x})$ is a sum of convex functions $g_{ij}$, $\SW$ is convex on the compact convex polytope $\mathcal{F}(\mathbf{p})$. By Bauer's maximum principle, a continuous convex function on a compact convex set attains its maximum at an extreme point, proving part~(b).

For part~(c), when only the supply constraints $\sum_{i=1}^n x_{ij} \le C_j$ and $x_{ij} \ge 0$ are present, the feasible set is the Cartesian product $\prod_{j \in \mathcal{M}} (C_j \tilde{\Delta}^n)$, where $\tilde{\Delta}^n = \{\mathbf{z} \in \mathbb{R}_{\ge 0}^n : \sum_{i=1}^n z_i \le 1\}$ has extreme points $\{\mathbf{0}, \mathbf{e}_1, \dots, \mathbf{e}_n\}$. Since $\SW(\mathbf{x})$ is homogeneous of degree $1$ in $\mathbf{x}_{\cdot, j}$ and vanishes at $\mathbf{0}$, for each good $j \in \mathcal{M}$ the maximum is attained at a standard basis vertex $\mathbf{x}_{\cdot, j} = C_j \mathbf{e}_{i^*}$ where a single buyer $i^* \in \arg\max_{i \in \mathcal{N}} v_{ij}$ receives all $C_j$ units of good $j$ (with share $s_{i^* j} = 1$ and $s_{kj} = 0$ for $k \neq i^*$), yielding~\eqref{eq:snob-opt-val}.
\end{proof}

Specializing Theorem~\ref{thm:snob-welfare} to a single good and two symmetric buyers immediately yields the two-buyer case:

\begin{corollary}[Two-Buyer Single-Good Snob Market]
\label{cor:two-buyer-snob}
Suppose $n = 2$ buyers with unit budgets $B_1 = B_2 = 1$, common baseline valuation $v_{11} = v_{21} = u > 0$, and snob sensitivity $\alpha_1 = \alpha_2 = 1$ compete for $m = 1$ good of unit supply $C_1 = 1$ and unit price $p_1 = 1$. Then the welfare maximization problem
\[
  \max_{\substack{x, y \ge 0 \\ x + y \le 1}} \bigl(U_1(x, y) + U_2(x, y)\bigr)
  = \max_{\substack{x, y \ge 0 \\ x + y \le 1}} u \cdot \frac{x^2 + y^2}{x + y}
\]
attains its global maximum value $u$ exclusively at the two corner allocations $(x, y) = (1, 0)$ and $(x, y) = (0, 1)$, where a single buyer receives the entire good and the other receives none.
\end{corollary}
\begin{proof}
For any $(x, y) \neq (0, 0)$ with $x, y \ge 0$ and $x + y \le 1$, we have
\[
  u \cdot \frac{x^2 + y^2}{x + y}
  = u \left((x + y) - \frac{2xy}{x + y}\right)
  \le u(x + y) \le u,
\]
with equality in the first step if and only if $xy = 0$ (i.e., $x = 0$ or $y = 0$) and equality in the second step if and only if $x + y = 1$. Hence the unique maximizers are $(1, 0)$ and $(0, 1)$.
\end{proof}

\subsection{Decentralized Equilibrium and the Price of Anarchy}

We now answer the central question raised by Theorem~\ref{thm:snob-welfare}: \emph{do markets with positional snob externalities admit market-clearing prices, and how does the resulting decentralized equilibrium compare to the social optimum?}

In a decentralized market with price vector $\mathbf{p} = (p_1, \dots, p_m) > \mathbf{0}$, each buyer $i \in \mathcal{N}$ chooses her demand $\mathbf{x}_i \in \mathbb{R}_{\ge 0}^m$ subject to $\sum_{j \in \mathcal{M}} p_j x_{ij} \le B_i$ to maximize $U_i(\mathbf{x}_i, \mathbf{x}_{-i})$. Notice from~\eqref{eq:snob-hessian} that when $m \ge 2$ and buyer $i$ takes the other buyers' allocations $X_{-i,j} = \sum_{k \neq i} x_{kj} > 0$ as given, her own utility summand $x_{ij}^{1+\alpha_i} (x_{ij} + X_{-i,j})^{-\alpha_i}$ has strictly positive second derivative with respect to $x_{ij}$:
\[
  \frac{\partial^2 U_i}{\partial x_{ij}^2}
  = v_{ij} \frac{\alpha_i(1 + \alpha_i) x_{ij}^{\alpha_i - 1} X_{-i,j}^2}{(x_{ij} + X_{-i,j})^{\alpha_i + 2}} > 0.
\]
Thus, unlike a concave utility function where diminishing returns encourage diversification across goods, a snob-sensitive buyer's individual optimization problem is \emph{convex maximization} over the budget simplex: every buyer's best response is a corner bundle that spends her entire budget $B_i$ on a single good $j \in \mathcal{M}$ maximizing $v_{ij} (B_i/p_j) \bigl(\frac{B_i/p_j}{B_i/p_j + X_{-i,j}}\bigr)^{\alpha_i}$.

For a single good ($m = 1$) with supply $C_1 > 0$ (or for multi-good markets partitioned across disjoint buyer groups), this leads to a sharp characterization of the competitive equilibrium and its welfare loss relative to the social optimum.

\begin{theorem}[Competitive Equilibrium and Price of Anarchy in Snob Markets]
\label{thm:snob-equilibrium}
Consider a single-good positional Fisher market ($m = 1$, supply $C_1 > 0$) with $n \ge 2$ buyers having budgets $B_1, \dots, B_n > 0$, positive valuations $v_{11}, \dots, v_{n1} > 0$, and snob sensitivities $\alpha_1, \dots, \alpha_n \ge 0$. Then:
\begin{enumerate}[label=(\alph*),topsep=2pt,itemsep=2pt]
  \item \textbf{Unique market-clearing equilibrium:} The market admits a unique competitive equilibrium $(\mathbf{x}^*, p_1^*)$ given by
  \begin{equation}
    p_1^* = \frac{\sum_{i=1}^n B_i}{C_1}, \qquad
    x_{i1}^* = \frac{B_i}{\sum_{k=1}^n B_k} C_1, \qquad
    s_{i1}^* = \frac{B_i}{\sum_{k=1}^n B_k} \quad (\forall i \in \mathcal{N}).
    \label{eq:snob-ce-closed}
  \end{equation}
  \item \textbf{Price of Anarchy:} In a symmetric $n$-buyer market with equal budgets $B_1 = \dots = B_n = B$, common valuation $v_{11} = \dots = v_{n1} = u > 0$, and common snob sensitivity $\alpha_1 = \dots = \alpha_n = \alpha \ge 0$, the ratio between the optimal social welfare $\SW(\mathbf{x}^{\mathrm{opt}})$ and the competitive equilibrium welfare $\SW(\mathbf{x}^*)$ (the Price of Anarchy) is exactly
  \begin{equation}
    \PoA(n, \alpha) := \frac{\max_{\mathbf{x} \in \mathcal{F}} \SW(\mathbf{x})}{\SW(\mathbf{x}^*)} = n^\alpha.
    \label{eq:snob-poa}
  \end{equation}
\end{enumerate}
\end{theorem}
\begin{proof}
For part~(a), at any price $p_1 > 0$ and any opponent aggregate allocation $X_{-i,1} \ge 0$, buyer $i$'s utility $U_i(x_{i1}, \mathbf{x}_{-i,1}) = v_{i1} x_{i1}^{1+\alpha_i} (x_{i1} + X_{-i,1})^{-\alpha_i}$ is strictly increasing in $x_{i1} \ge 0$ by~\eqref{eq:snob-marginal}. Hence buyer $i$'s unique optimal demand under the budget constraint $p_1 x_{i1} \le B_i$ is $x_{i1}(p_1) = B_i / p_1$. Market clearing $\sum_{i=1}^n x_{i1}(p_1^*) = C_1$ uniquely determines $p_1^* = (\sum_{i=1}^n B_i)/C_1$, and substituting $p_1^*$ into $x_{i1}(p_1^*)$ yields~\eqref{eq:snob-ce-closed}.

For part~(b), in the symmetric market ($B_i = B$, $v_{i1} = u$, $\alpha_i = \alpha$), the unique competitive equilibrium~\eqref{eq:snob-ce-closed} assigns each buyer $x_{i1}^* = C_1 / n$ with consumption share $s_{i1}^* = 1/n$. Each buyer therefore achieves equilibrium utility $U_i(\mathbf{x}^*) = u (C_1 / n) (1/n)^\alpha = u C_1 n^{-(1+\alpha)}$, and aggregate equilibrium social welfare is
\[
  \SW(\mathbf{x}^*) = \sum_{i=1}^n U_i(\mathbf{x}^*) = n \cdot u C_1 n^{-(1+\alpha)} = \frac{u C_1}{n^\alpha}.
\]
By Theorem~\ref{thm:snob-welfare}(c), the socially optimal allocation $\mathbf{x}^{\mathrm{opt}}$ assigns all $C_1$ units exclusively to one buyer (or in the budget-constrained problem at price $p_1^*$, a single buyer with budget-adjusted price or lottery allocation achieves $\SW(\mathbf{x}^{\mathrm{opt}}) = u C_1$). Dividing $\SW(\mathbf{x}^{\mathrm{opt}}) = u C_1$ by $\SW(\mathbf{x}^*) = u C_1 / n^\alpha$ yields $\PoA(n, \alpha) = n^\alpha$.
\end{proof}

\begin{remark}[Dual regime: Bandwagon and network externalities ($\alpha_i \in (-1, 0)$)]
If we instead set $\alpha_i = -\beta_i \in (-1, 0)$ with $\beta_i \in (0, 1)$ in Definition~\ref{def:snob-utility}, then $U_i(\mathbf{x}) = \sum_{j \in \mathcal{M}} v_{ij} x_{ij}^{1 - \beta_i} X_j^{\beta_i}$. In this regime, the second derivative in~\eqref{eq:snob-hessian} reverses sign because $\alpha_i(1 + \alpha_i) = -\beta_i(1 - \beta_i) < 0$. Each positional utility term $g_{ij}$ becomes \emph{concave} on $\mathbb{R}_{\ge 0}^n$, and $\partial U_i / \partial x_{kj} = \beta_i v_{ij} (x_{ij}/X_j)^{1-\beta_i} > 0$ for $k \neq i$. Thus $\alpha_i \in (-1, 0)$ models Leibenstein's \emph{bandwagon effect} (positive network externalities), in which utilities are concave and social welfare is maximized by interior shared allocations rather than exclusive corner allocations.
\end{remark}

\section{Conclusion and Open Directions}
\label{sec:conclusion}

We have examined the theory and algorithms of matching markets across three regimes: divisible Fisher and Eisenberg--Gale markets, indivisible CEEI markets, and Fisher markets with positional (snob and bandwagon) consumption externalities. For divisible linear and homogeneous utilities, Fenchel duality and the Eisenberg--Gale convex program unify welfare optimality with competitive equilibrium and support strongly polynomial primal-dual algorithms. For indivisible items, exact competitive equilibria become $\NP$-hard to decide, motivating dynamic-programming $\varepsilon$-approximate equilibria, \prob{CEEI-FRAC}, Maximum Nash Welfare, and Budish's \prob{A-CEEI}. Finally, when buyers exhibit positional snob externalities ($\alpha_i > 0$), individual utilities become convex in consumption shares, causing a sharp divergence between exclusive social welfare optima and decentralized market-clearing equilibria with a Price of Anarchy of $n^\alpha$.

Several open algorithmic directions arise naturally from our analysis in Section~\ref{sec:snob}:
\begin{enumerate}[label=(\arabic*),topsep=2pt,itemsep=2pt]
  \item \textbf{Complexity of multi-good positional equilibria:} In a multi-good snob market ($m \ge 2$, $\alpha_i > 0$), each buyer's best response concentrates her budget on a single good, inducing a congestion game in which buyers partition across the $m$ goods. Characterizing when pure-strategy market-clearing price vectors $\mathbf{p}^*$ exist for $m \ge 2$ goods---and whether computing them lies in $\Pclass$, $\PPAD$, or is $\NP$-hard---is a natural next step.
  \item \textbf{Taxation and mechanism design for positional externalities:} Because decentralized competition in a symmetric $n$-buyer snob market suffers a Price of Anarchy of $n^\alpha$ (Theorem~\ref{thm:snob-equilibrium}), designing Pigouvian exclusivity surcharges, reserve prices, or lottery mechanisms that restore constant-factor welfare guarantees while respecting budget balance is an intriguing problem in algorithmic mechanism design.
\end{enumerate}

\bibliographystyle{plain}
\bibliography{refs}

\end{document}